\documentstyle[sprocl,epsfig]{article}

\begin{document}

\title{ENHANCEMENT OF COULOMB DRAG AWAY FROM HALF FILLED LANDAU LEVELS}

\author{M. P. Lilly,$^1$ J. P. Eisenstein,$^1$ L. N. Pfeiffer,$^2$ and K. W. West$^2$}

\address{$^1$Condensed Matter Physics, Caltech, Pasadena, CA 91125\\ 
         $^2$Bell Laboratories, Lucent Technologies, Murray Hill, NJ 07974}


\maketitle\abstracts{Coulomb drag between two parallel two dimensional
electron gases has been measured at high magnetic fields in samples with
two different layer spacings. As the layer filling factor deviates from
$\nu=1/2$, we find that the magnitude of the drag is enhanced
quadratically with $\Delta \nu = \nu -1/2$, and the
curvature of the enhancement is insensitive to both the sign of $\Delta \nu$ 
and the spacing between the layers. Our results suggest
that the enhancement is {\it not} due to non-perturbative interlayer
correlations.}

Double layer two-dimensional electron gases (2DEGs) 
have been of recent interest due to the
rich many-body physics they exhibit. Of particular interest
is the case when each of the 2DEGs is
at filling factor $\nu_{tot} = 1/2 + 1/2$. If the separation between the
layers is large relative to the interparticle separation, the system
behaves as two individual layers, each of which is widely viewed at
a composite fermion (CF) liquid. If the separation is small, a gap
develops and the system forms a ferromagnetic quantum Hall state
characterized by total filling factor $\nu_{tot} = 1$. The
nature of this transition is a frontier topic in the
field.\cite{perspectives}

Novel transport techniques such
as Coulomb drag can be utilized in double layer 2DEGs. 
In a Coulomb drag measurement, current flowing in one
layer induces a voltage in the other whose magnitude measures the
interlayer momentum relaxation rate, $\tau_m$. At high magnetic fields
$\tau_m$ in our samples is dominated by electron-electron scattering
between the layers.
Previous measurements of Coulomb drag at half-filling\cite{lilly} 
show the drag to be typically 1000 times
larger than at B=0.  Models of weakly coupled CF layers\cite{kim,stern,sakhi} 
also predict this large increase of the drag and attribute it to the
increased
importance of electron interactions in the lowest Landau level.
In this paper we report measurements of
Coulomb drag between 2DEGs where the layer filling factor is
systematically varied around $\nu = 1/2$.  Two samples having different layer 
separations have been studied.  We find that
the drag increases quadratically as  a function of $\Delta \nu = \nu - 1/2$,
where $\nu = hn/eB$, $n$ is the density, and $B$ is the magnetic field.

The two samples studied here are modulation doped GaAs/AlGaAs double
quantum wells grown by molecular beam epitaxy. They consist of two
200~\AA\ wide GaAs wells separated by an Al$_x$Ga$_{1-x}$As barrier.
Sample A has a 100~\AA\ barrier with $x=1$ and sample B has a 225~\AA\
barrier with $x=0.32$. The electron densities in both samples A and B
were balanced using central top and bottom gates to a nominal density of
$n \approx 1.4 \times 10^{11}$~cm$^{-2}$. Each sample was patterned into a
Hall bar ($l = 400 \mu m$, $w = 40 \mu m$) with indium ohmic contacts. Drag
measurements are made by injecting a current, $I$ = 10 nA, 13 Hz in one 2DEG
(drive layer) and measuring resulting voltage, $V_D$, induced in the other (drag
layer). 

\begin{figure}[t]
\begin{center}
\epsfig{file=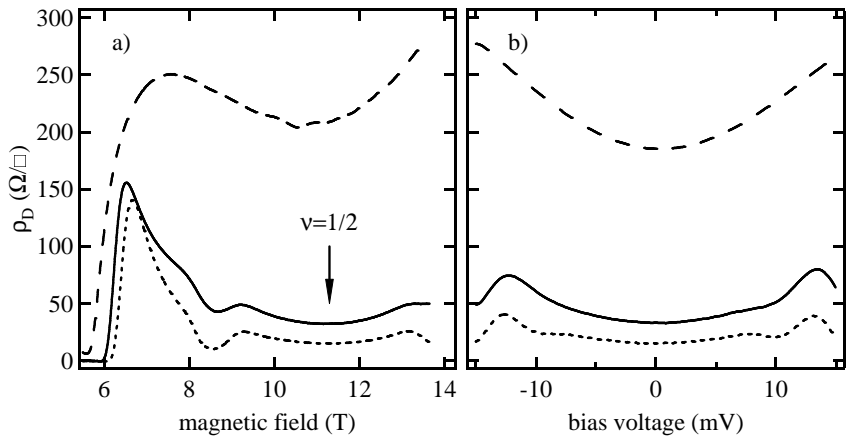,bbllx=151pt,%
        bblly=464pt,bburx=504pt,bbury=648pt,width=4.5in}
\end{center}
\caption[fig1]{In (a) the magnetic field dependence of drag in sample A
is shown for $T = 0.6$~K (dotted), 1.0~K (solid), and 4.0~K (dashed). In (b) an
interlayer bias voltage is applied to sample A at a constant magnetic
field for 0.6~K (dotted), 1.0~K(solid), and 3.5~K (dashed).} 
\end{figure}

Fig. 1a shows the drag resistivity, $\rho_D = -(w/l)V_D/I$, of sample A for 
$\nu \le 1$ at several temperatures. The sign of the drag voltage is
opposite that of the resistive voltage drop in the drive layer. This sign 
is consistent with electrons in the drag layer building up a voltage to oppose the momemtum
transfer of the electrons from the drive layer. At $B = 5.7$~T, each
layer is in the integer quantum Hall state $\nu=1$. 
Owing to the large energy gaps in each layer at this filling, 
the drag $\rho_D \rightarrow 0$. 
At $T=0.6$~K, the same effect is evident as the fractional 
quantum Hall effect (FQHE) $\nu = 2/3$ develops at
$B=8.55$~T. Around $\nu=1/2$ in each layer, the 2DEGs remain
compressible, and $\rho_D$ forms a shallow minimum.
As the temperature is lowered, the magnitude of the
drag decreases, although the minimum around $\nu=1/2$ persists. 

A second measurement which yields enchanced drag away from
half-filling occurs when the densities of the layers are intentionally
unbalanced at fixed magnetic field. 
Starting at $\nu = 1/2$ with density $n_0$ in each layer, 
a dc bias voltage, $V_{bias}$ is applied between the
layers. The bias voltage transfers charge between layers, 
increasing the density of one layer to $n_0 + \Delta n/2$ and 
decreasing it in the other to $n_0 - \Delta n/2$ ($\Delta n$ 
is the density difference).  The change in density
at a fixed field causes a corresponding change in the filling factor that 
we define by $\Delta \nu = \nu - 1/2$.
The results for sample A
are shown in Fig. 1b at T=0.6~K, 1.0~K and 3.5~K. 
At $V_{bias} = 0$, both layers are at $\nu=1/2$. For
small $V_{bias}$, $\rho_D$ increases
quadratically with a curvature that depends on temperature. The minimum in
$\rho_D$ in the $T=0.6$~K data (dotted line) at $V_{bias} = 10$~mV is due to
the developement the FQHE at $\nu_{tot} = 2/5 + 3/5$. This feature allows
calibration between $V_{bias}$ and $\Delta \nu$.  A similar
increase in $\rho_D$ as a function of $V_{bias}$ was found in sample B.

\begin{figure}[t]
\begin{center}
\epsfig{file=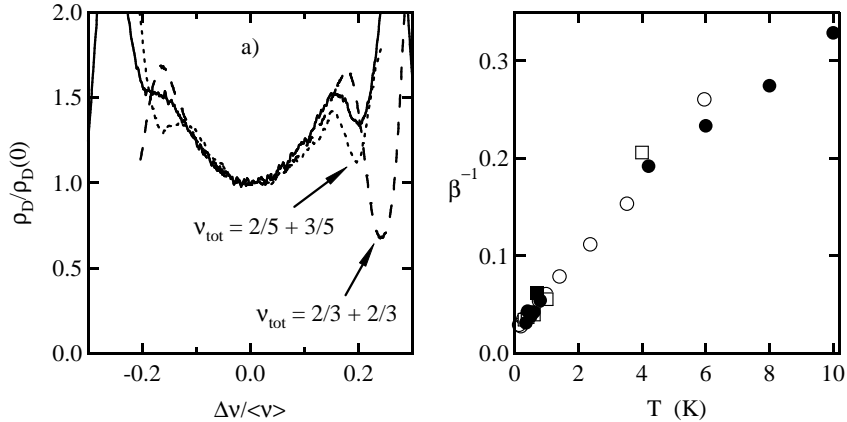,bbllx=144pt,%
        bblly=464pt,bburx=508pt,bbury=652pt,width=4.5in}
\end{center}
\caption[fig2]{(a) Scaled drag resistivity as filling factor departs
from $\nu=1/2$ at T=0.6 K for sample A field sweep (dashed), sample A 
interlayer bias (solid) and sample B interlayer bias (dotted).
(b) Summary of curvature at different temperatures for sample A field sweep
(open square), sample A interlayer bias (open circle), sample B field sweep
(solid square), and sample B interlayer bias (solid circle).}
\end{figure}

In order to compare the enhancement of the drag away from
half-filling for the two situations described above, 
we plot the normalized drag resistivity, $\rho_D/\rho_D(\Delta\nu=0)$, 
as a function of the relative change in 
filling factor, $\Delta\nu / \langle \nu \rangle$, where $\Delta \nu = 
\nu - 1/2$ and $\langle \nu \rangle = (\nu_1+\nu_2)/2$ (the subscripts indicate
the layer index).  When $V_{bias}$ is
applied, $\Delta \nu_1 = - \Delta \nu_2$, and $\langle \nu \rangle = 1/2$ is 
constant with $V_{bias}$.  When the field is changed, 
$\Delta \nu_1 = \Delta \nu_2$, and $\langle \nu \rangle = h n / B$  changes
with the field.  
Results for $T=0.6$ K are shown in Fig. 2a for interlayer bias in sample A
(solid line), magnetic field for sample A (dashed) and interlayer bias for sample B
(dotted).  
For small $\Delta \nu$, the relative drag signal increases quadratically for a
with the same curvature for all three sets of data.
At larger $\Delta \nu$, features related to the FQHE are observed at $\Delta \nu / \langle \nu
\rangle = 0.2$ ($\nu_{tot} = 2/5 + 3/5$) in the interlayer bias traces (solid and dotted
lines) and at
$\Delta \nu / \langle \nu \rangle = 0.25$ ($\nu_{tot} = 2/3 + 2/3$) for the field sweep
(dashed line).

Defining the curvature by $\rho_D/\rho_D(0) = \beta
(\Delta\nu/\langle\nu\rangle)^2 + 1$,
in Fig. 2b we show the temperature dependence of $\beta^{-1}$.
The open symbols are results for sample A (squares for field
sweeps and circles for interlayer bias) and the closed symbols for
sample B.  Extrapolating to $T=0$, $\beta^{-1}$ appears to have a finite
intercept.
We note that $\beta$ follows the same curve in all cases.  
Comparing the
two different measurements on the same sample, it may not be suprising
that the enhancement of the drag should not depend on  
whether the field is changed
or an interlayer bias is applied, especially for small changes in $\Delta \nu$.
On the other hand, comparing the curvatures of sample A to sample B
which have different layer separations, the data also fall on
the same curve.  The enhancement of the drag does not depend on the 
spacing between the 2D layers.  From this independence, we conclude that
the enhancement in drag is due to the response of a single layer, and {\it not}
due to interlayer correlations.

In conclusion, measurements of Coulomb drag near half filling show a
quadratic increase in the drag as a function of $\Delta \nu$.  The 
increase in the drag signal for a given sample does not depend on whether
the density is changed or whether the magnetic field is changed, and is
symmetric around $\nu=1/2$.  In addition, the
curvature does not depend on the spacing between the layers.  This
suggests that the enhanced drag signal is due to the response of the 
individual layers.

We acknowledge useful discussions with A. Stern.  This work was supported by
the NSF via DMR-9700945.

\end{document}